\RequirePackage{etoolbox}
\csdef{input@path}{{style/}{graphics/}}
\documentclass[ba,linksfromyear,nameyear,dvips,preprint]{imsart}
\usepackage{natbib}
\usepackage[backref=page]{hyperref}
\usepackage{amsthm,amsmath}
\usepackage{graphicx}

\usepackage{algorithm}
\usepackage{algorithmic}
\usepackage{xcolor}

\pubyear{2015}
\volume{10}
\issue{1}
\firstpage{171}
\lastpage{187}
\doi{10.1214/14-BA891}

\startlocaldefs
\def \xo {x_{obs}}
\def \xti {\tilde{x}}
\def \theti {\tilde{\theta}}
\endlocaldefs

\begin{document}

\begin{frontmatter}
\title{Sequential Monte Carlo with Adaptive Weights for Approximate Bayesian Computation}
\runtitle{Adaptive Weight ABC SMC}

\begin{aug}
\author[a,b]{\fnms{Fernando V.} \snm{Bonassi}\corref{}\ead[label=e1]{bonassi@gmail.com}}
\and
\author[c]{\fnms{Mike} \snm{West}\ead[label=e2]{mw@stat.duke.edu}}

\runauthor{F. V. Bonassi and M. West}

\address[a]{Google Inc, Mountain View, CA, \printead{e1}}
\address[b]{Research performed as a PhD student in Statistical Science at Duke University}
\address[c]{Duke University, Durham, NC, \printead{e2}}
\end{aug}

\begin{abstract}
Methods of approximate Bayesian computation (ABC) are increasingly used for analysis of complex models.
A major challenge for ABC is over-coming the often inherent problem of high rejection rates in the accept/reject methods
based on prior:predictive sampling. A number of recent developments aim to address this with extensions
based on sequential Monte Carlo (SMC) strategies. We build on this here, introducing an ABC SMC method
that uses data-based adaptive weights. This easily implemented and computationally trivial extension of ABC SMC
can very substantially improve acceptance rates, as is demonstrated in a series of examples with simulated and
real data sets, including a currently topical example from dynamic modelling in systems biology applications.
\end{abstract}


\end{frontmatter}


\section{Introduction}
\label{sec:intro}

Methods of approximate Bayesian computation (ABC)  are becoming increasing exploited, especially
for problems in which the likelihood function is analytically intractable or very expensive to compute~\citep{Pritchard1999,Marjoram2003}. Recent applications run across  areas such as evolutionary genetics~\citep{Beaumont2002}, epidemiology~\citep{McKinley2009}, astronomical model analysis~\citep{Cameron2012}, among others~\citep{Csillery2010}.

Vanilla ABC simulates parameter/data pairs $(\theta,x)$ from the prior distribution whose density is $p(\theta)p(x|\theta),$
accepting $\theta$ as an approximate posterior draw if its companion data $x$ is \lq\lq close enough'' to the observed data $\xo.$
If $\rho(x,\xo)$ is the chosen measure of discrepancy, and $\epsilon$ is a discrepancy threshold defining \lq\lq close'', then accepted
parameters are a sample from  $p(\theta | \rho (x, \xo) < \epsilon).$
Often, it is not possible or efficient to use the complete data set. In that case, a reduced
dimensional set of summary statistics $S$ can be used so that the accepted draws sample from $p(\theta | \rho (S,S_{obs}) < \epsilon).$ For simplicity
of notation, the posterior will be denoted by $p(\theta | \rho (x, x_{obs}) < \epsilon)$, with the understanding that  the
effective data $x$ potentially represents such a summary of the original data.

A main issue is high rejection rates that result from:  ($i$) the requirement that $\epsilon$ be small to build faith that the approximate posterior is a good
approximation to $p(\theta|\xo), $ and  ($ii$) the posterior $p(\theta|\xo)$ may be concentrated in completely different regions of parameter space than the prior. To address this, modifications of vanilla ABC are emerging, including regression adjustment strategies~\citep[e.g.,][]{Beaumont2002,Blum2010,Bonassi2011}, and automatic sampling schemes~\citep{Marjoram2003,Sisson2007} utilizing techniques such as sequential Monte Carlo (SMC).
In the latter ABC methods, SMC is used in order to automatically, sequentially  refine posterior approximations to be used to generate proposals for further steps.  At each of a series of sequential steps indexed by $t,$ these methods aim to generate draws from $p(\theta | \rho (x, \xo) < \epsilon_t) $ where $\epsilon_t$ define a series of decreasing thresholds. Total acceptance rates can be significantly higher than with vanilla ABC scheme~\citep{Toni2009,Stumpf2010}.

The original version of the ABC SMC algorithm proposed in \cite{Sisson2007} was motivated by the SMC samplers methodology of~\cite{Delmoral2006}.
Later, \cite{Beaumont2009} realized that this original method can result in biased samples relative to the true posterior,
and this was followed by development of a corrected approach~\citep{Beaumont2009,Sisson2009,Toni2009}. The general form of the algorithm, which relies fundamentally on sequential importance sampling,  is shown in Figure~\ref{fig:ABC_SMC}.

Applications of this ABC SMC algorithm have been presented in a variety of areas including population genetics~\citep{Beaumont2009}, systems biology~\citep{Toni2009,Stumpf2010,Liepe2010} and psychology~\citep{Turner2012}.
In terms of methodology development, there is increasing interest in extensions and improvements of this algorithm, as demonstrated
in, for example, \cite{Lenormand2012,Filippi2012,Silk2012}.  Building on this momentum and open challenges to improving the
methodology, our current focus is on the form of the ABC SMC
of Figure~\ref{fig:ABC_SMC}.
We note and comment on the  ABC SMC approach of
\cite{Delmoral2011}; this makes use of an adaptive threshold schedule, extends SMC samplers~\citep{Delmoral2006} and uses
an Markov chain Monte Carlo (MCMC) kernel for propagation of particles.
As a result, the computation of weights has linear complexity as a function of the number of particles, while a disadvantage of this approach is that it can result in particle duplications, possibly leading to inferior overall performance \citep[as empirically shown in][]{Lenormand2012} in comparison to the basic algorithm of Figure~\ref{fig:ABC_SMC} that we begin with here.

One aspect of the algorithm of Figure~\ref{fig:ABC_SMC} is that the computational demand in evaluating weights increases quadratically as a function of the number of particles. While this appears disadvantageous,
in practice it is often just not a limiting factor.  The reason for this is that in practical
ABC applications, computation time is typically substantially dominated by the repeated data simulation steps; this
is borne out in our own experiences discussed further below, and has been clearly
indicated and discussed in other studies, including \cite{Beaumont2009,Filippi2012,Delmoral2011}, for example.

\begin{figure}[h!]
\noindent\begin{minipage}[c]{1\linewidth}
\begin{center}
\noindent \begin{minipage}[c]{0.85\linewidth}
\begin{center} \line(1,0){320} \end{center}
    \begin{enumerate}
     \item Initialize threshold schedule $\epsilon_1>\cdots>\epsilon_T$
    \item Set $t=1$ \\
    \hspace*{0.2cm} For $i=1,\ldots,N$ \\
    \hspace*{0.5cm} -- Simulate $\theta_i^{(1)} \sim p(\theta)$ and $x \sim p(x|\theta_i^{(1)})$ until $\rho(x,x_{obs}) < \epsilon_1$\\
    \hspace*{0.5cm} -- Set $w_i = 1/N$
    \vspace*{0.01cm}
    \item For $t=2,\ldots,T$ \\
    \hspace*{0.3cm} For $i=1,\ldots,N$ \\
    \hspace*{0.5cm} -- Repeat: \\
    \hspace*{1.5cm} Pick $\theta_i^{*}$ from the $\theta_j^{(t-1)}$'s with probabilities $w_j^{(t-1)}$, \\
    \hspace*{1.5cm} draw $\theta_i^{(t)} \sim K_t(\theta_i^{(t)}| \theta_i^{*})$ and $x \sim p(x|\theta_i^{(t)})$; \\
    \hspace*{0.5cm} \ \ until $\rho(x,x_{obs}) < \epsilon_t$ \vspace*{0.1cm} \\
    \hspace*{0.5cm} -- Compute new weights as  $$w_i^{(t)} \propto \frac{p(\theta_i^{(t)})}{\sum_j w_j^{(t-1)}K_t(\theta_i^{(t)}| \theta_j^{(t-1)})}$$
    \hspace*{0.3cm} Normalize $w_i^{(t)}$ over $i=1,\ldots,N$
     \end{enumerate}
 \end{minipage}
 \end{center}
 \end{minipage}
 \caption{ABC SMC algorithm~\citep{Beaumont2009,Sisson2009,Toni2009}. Here $K_t(\cdot|\cdot)$ is a conditional density that serves as
 a transition kernel to \lq\lq move'' sampled parameters and then appropriately weight accepted values. In contexts of real-valued parameters,
 for example,  $K_t(\theta|\theta^*)$ might be taken as a multivariate normal or $t$ density centred at or near  $\theta^*,$ and whose scales
 may decrease as $t$ increases.}
 \label{fig:ABC_SMC}
 \vspace*{6pt}
 \end{figure}

 \section{ABC SMC with Adaptive Weights \label{sec:AW} }

The ABC SMC strategy  above is intimately related to adaptive importance sampling \citep{Liu2001MonteCarlo,Cornuet2012} using adaptively refined posterior approximations based on
kernel mixtures as importance samplers. Originating as a method for direct posterior approximation~\citep{West1992b,West1993a}
in complex models, that approach defines kernel density representations of a \lq\lq current'' posterior approximation
$g_t(\theta) =\sum_j w_j K_t(\theta | \theta_j)$ as an importance sampler for a next step $t+1$, then adaptively updates the
parameters defining $K_t(\cdot|\cdot)$ as well as importance weights~\citep{West1993b,Liu2001}.

 \begin{figure}[htb!]
\noindent\begin{minipage}[c]{1\linewidth}
\begin{center}
\noindent \begin{minipage}[c]{0.85\linewidth}
\begin{center} \line(1,0){320} \end{center}
    \begin{enumerate}
     \item Initialize threshold schedule $\epsilon_1>\cdots>\epsilon_T$
    \item Set $t=1$ \\
    \hspace*{0.2cm} For $i=1,\ldots,N$ \\
    \hspace*{0.5cm} -- Simulate $\theta_i^{(1)} \sim p(\theta)$ and $x \sim p(x|\theta_i^{(1)})$ until $\rho(x,x_{obs}) < \epsilon_1$\\
    \hspace*{0.5cm} -- Set $w_i = 1/N$
    \item $t=2,\ldots,T$ \\
    \hspace*{0.2cm} Compute data based weights $v_{i}^{(t-1)} \propto w_i^{(t-1)} K_{x,t}(x_{obs}| x_i^{(t-1)})$\\
    \hspace*{0.2cm} Normalize  weights $v_{i}^{(t-1)}$ over $i=1,\ldots,N$ \vspace*{0.3cm} \\
    \hspace*{0.2cm} For $i=1,\ldots,N$ \\
    \hspace*{0.5cm} -- Repeat: \\
    \hspace*{1.5cm} Pick $\theta_i^{*}$ from the $\theta_j^{(t-1)}$'s with probabilities $v_j^{(t-1)}$, \\
    \hspace*{1.5cm} draw $\theta_i^{(t)} \sim K_{\theta,t}(\theta_i^{(t)}| \theta_i^{*})$ and $x \sim p(x|\theta_i^{(t)})$; \\
    \hspace*{0.5cm} \ \ until $\rho(x,x_{obs}) < \epsilon_t$ \vspace*{0.1cm} \\
    \hspace*{0.5cm} -- Compute new weights as
             $$w_i^{(t)} \propto \frac{p(\theta_i^{(t)})}{\sum_j v_j^{(t-1)}K_{\theta,t}(\theta_i^{(t)}| \theta_j^{(t-1)})}$$
                \hspace*{0.3cm} Normalize $w_i^{(t)}$ over $i=1,\ldots,N$
     \end{enumerate}
 \end{minipage}
 \end{center}
 \end{minipage}
 \caption{ABC SMC with Adaptive Weights.}
 \label{fig:ABC_SMC_AW}
 \end{figure}

This historical connection motivates an extension of ABC SMC that is the focus of this paper. That is, simply
apply the idea of kernel density representation to the joint distribution of accepted values $(x,\theta)$ using a joint kernel
$K_t(x, \theta| x^*, \theta^*).$  For this paper, we use a product kernel $K_t(x, \theta| x^*, \theta^*) = K_{x,t}(x | x^*) K_{\theta, t} (\theta | \theta^*)$, which
leads to major benefits in terms of computational convenience. The underlying idea that we are working with  kernel density
approximations to the joint distribution of $(x,\theta)$ means that we can rely on the utility of product kernel mixtures generally~\citep[e.g.,][]{Fryer1977,Scott2005}.  A joint approximation $g_t(\theta,x) \propto \sum_j w_j K_{x,t}(x | x_j) K_{\theta,t} (\theta | \theta_j)$
yields a marginal mixture for each of $x$ and $\theta$ separately, and a posterior approximation (emulator or importance sampler)
with density $g_t(\theta | \xo) \propto \sum_j w_j K_{x,t}(\xo | x_j) K_{\theta,t} (\theta | \theta_j)$. In our proposed extension of ABC SMC,
this form is used  to  propose new values, and we can immediately see how proximity of any one $x_i$ to the observed data $\xo$ will now
help raise the importance of proposals drawn from or near to the partner particle $\theta_i. $
As for $K_{\theta,t}(\cdot | \theta)$, multivariate normal or $t$ density are natural choices for $K_{x,t}(\cdot | x)$. For a more detailed discussion of kernel choices, see~\cite{Silverman1986} and \cite{Scott1992}.

Figure~\ref{fig:ABC_SMC_AW} shows the algorithmic description of this new ABC SMC with Adaptive Weights (ABC SMC AW).
The inclusion of a new step where the weights are modified according to the respective values of $x$ adds computations. However,
since computational time in the ABC SMC algorithm is usually dominated by the extensive repetition of model simulations, the
increased compute burden will often be negligible.
Note also that the original ABC SMC is a particular case when $K_{x,t}(\cdot | x)$ is uniform over the region of accepted values of $x$.

The idea of approximating the joint distribution of parameters and data together is also present in \cite{Bonassi2011}, where mixture modeling is applied to the joint distribution and then used as a form of nonlinear regression adjustment. The smoothing step in ABC SMC AW can  be seen as an automatic simplified version of that approach, about which we say more in the concluding Section~\ref{sec:Additional Discussion}.

\section{Theoretical Aspects}

We discuss some of the structure of ABC SMC AW to provide insight as to why it can be expected to
lead to improved acceptance rates.

First, note that the final output of ABC SMC methods is an estimate of the target distribution $p (\theta | d(x,\xo) < \epsilon_T)$.
This provides the practitioner with the convenient option of using the output as a final approximation or as an input to be refined by using a preferred regression adjustment technique, such as local linear regression~\citep{Beaumont2002} or mixture modeling~\citep{Bonassi2011}. However, in the context of intermediate ABC SMC steps, more refined approximations have the potential to improve efficiency; this can be automatically
achieved by use of kernel smoothing techniques. At each intermediate step $t$, this motivates the following approximation for the joint
density $p(x,\theta$), locally around $x=\xo$:
\begin{eqnarray}
\hat{p}_t (x,\theta | A) = \int  \int p ( \theti | A) p ( \xti | \theti, A) K_{\theta} ( \theta | \theti ) K_x( x | \xti ) d \xti d\theti
 \label{eq:ABC_2}
\end{eqnarray}
where $A = \{x: d(x,\xo) < \epsilon_t \}$ represents the acceptance event at step $t$, $K_{\theta}$ and $K_{x}$ are the kernel functions as described in Section~\ref{sec:AW}, and approximate draws of $p(\theti, \xti | A)$ are obtained via importance sampling.

This defines an implicit posterior emulator, namely
\begin{eqnarray}
\hat{p}_t ( \theta | \xo ) \ \ \propto \ \ \int  \int p ( \theti | A) p ( \xti | \theti, A) K_{\theta} ( \theta | \theti ) K_x( \xo | \xti ) d \xti d\theti.
 \label{eq:ABC_3}
\end{eqnarray}
We now explore equation~(\ref{eq:ABC_3}) to ensure that: (A) it defines a valid importance sampler with
target $p(\theta | \rho(x,\xo) < \epsilon_t)$ at each step $t$,  and (B) the resulting expected acceptance rates are higher than for regular ABC SMC.
We address these two points in turn.

\paragraph{(A):}
At step $t,$ note that $\theta$ is sampled from the distribution having density
$g(\theta) = \sum_j v_j^{(t-1)}K_{t,\theta}(\theta | \theta_j^{(t-1)}).$   Given this
value,  the extra step then simulates $x$ until the event $A$ is true. This event has probability $pr (  A  |\  \theta)$, where $pr (A  |\  \theta) = \int_{A} p(x, \theta) dx$, resulting in the  overall proposal
\begin{eqnarray}
g(\theta | A) \ \ \propto \ \ g(\theta) pr (A  | \ \theta).
 \label{eq:proposal_distribution}
\end{eqnarray}
Finally, to sample from $p (\theta | A) = p (\theta | d(x,\xo) < \epsilon_t)$ based on this proposal, the importance sampling weights are
\begin{eqnarray}
w(\theta) \ \ \propto \ \ \frac{p(\theta) pr (A  | \ \theta)}{g(\theta) pr (A  | \ \theta)} = \frac{p(\theta)}{g(\theta)},
 \label{eq:proposal_weight}
\end{eqnarray}
which have exactly the same form as in Figure~\ref{fig:ABC_SMC_AW}.

\paragraph{(B):}
At step $t,$ equation~(\ref{eq:ABC_2}) describes the approximation for the joint density of $(x,\theta)$ locally around $x=\xo$.
Based on this representation, it is possible to show that the proposal distribution implicitly defined in ABC SMC AW results in higher
prior predictive density over the acceptance region for the next SMC step.
This is seen as follows.  For simplicity of notation, use $p_0(x,\theta)$  to refer to $\hat{p}_t (x, \theta)$ at the current step~$t$.
The proposal density in ABC SMC is $p_0(\theta)$, whereas that for ABC SMC AW is $p_0(\theta | \xo)$.
These two densities induce marginal prior predictive densities $p_0(x)$ and $p_1(x),$ respectively.
Integration of a prior predictive density over the acceptance region $A_{t+1} = \{x: \rho(x,\xo) < \epsilon_{t+1} \}$
yields the corresponding  acceptance probability for the next SMC step, namely  $pr (A_{t+1})$.
We now show that $pr_1(A_{t+1}) > pr_0(A_{t+1})$ so that ABC SMC AW improves acceptance rates over regular ABC SMC.
Our proof relies on the assumption that acceptance probability $pr_0(A_{t+1} | \Theta)$ is positively correlated with $p_0(\xo | \Theta)$ with respect to $\theta\sim p_0(\theta)$. This is a reasonable assumption to make in regular cases,
including the limiting case that $\epsilon_{t+1} \to 0$ when the correlation tends to 1.

Under ABC SMC we have
\begin{eqnarray}
pr_0(A_{t+1}) & = &\int pr_0(A_{t+1} | \theta) p_0( \theta) d\theta = E(pr_0(A_{t+1} | \Theta)),  \nonumber
\end{eqnarray}
where the expectation is with respect to $p_0(\cdot).$  The corresponding value under ABC SMC AW is
\begin{eqnarray}
pr_1(A_{t+1}) & = &\int pr_0(A_{t+1} | \theta) p_0( \theta | \xo) d\theta \nonumber  \\
          & =  & \int pr_0(A_{t+1} | \theta) \frac{p_0(\xo | \theta)p_0(\theta)}{p_0(\xo)} d\theta = \frac{E(pr_0(A_{t+1} | \Theta) p_0(\xo | \Theta))}{p_0(\xo)} \nonumber  \\
          & > &     \frac{E(pr_0(A_{t+1} | \Theta)) E(p_0(\xo | \Theta)) }{p_0(\xo)} = pr_0(A_{t+1}),
\nonumber
\end{eqnarray}
assuming $Cov(pr_0(A_{t+1} | \Theta) p_0(\xo | \Theta)) > 0$.

\section{Illustrative Example:  Normal Mixtures}

\subsection{A Standard Example} \label{sec:Normal_toy}

A  simple example  taken from previous studies of ABC SMC~\citep{Sisson2007,Beaumont2009} concerns scalar data  $x | \theta \sim 0.5 \mathcal{N}(\theta,1) + 0.5 \mathcal{N}(\theta,0.01)$ and prior  $\theta \sim \mathcal{U}(-10,10)$.
With observed value $\xo=0,$ the target posterior is $\theta | \xo \sim 0.5 \mathcal{N}(0,1) + 0.5 \mathcal{N}(0,1/100)$ truncated to $(-10,10)$.

We follow details in \cite{Sisson2007} with discrepancy measure $\rho(x,\xo)=|x-\xo|$ and threshold schedule $\epsilon_{1:3} = (2,0.5,0.025),$
and we use normal kernels $K_{\theta,t}$ and $K_{x,t}$ with standard deviations (or  bandwidth parameters) $h_{\theta}$ and $h_x,$
respectively.  Following standard recommendations in~\cite{West1993a} and \cite{Scott2005}, the bandwidths $h_k$, for $k\in \{ x,\theta \},$
are set at
 $h_k = \hat{\sigma}_k/N^{1/6}$  where $\hat{\sigma}_k$ is the standard deviation, which is computed based on the values of the particles and their respective weights.
This standard rule-of-thumb specification is an asymptotic approximation to the optimal bandwidth choice based on the mean integrated squared error of the product kernel
density estimate~\citep{Scott1992}.

\begin{table}[h!]
\begin{center}
  \begin{tabular}{ c c c c }
$t$     & $\epsilon_t $ & ABC SMC & ABC SMC AW  \\
1   & 2         & 5.01      & 4.96  \\
2   & 0.5   & 4.33      & 2.38  \\
3   & 0.025     & 39.71     & 27.22 \\
\hline
    & Total     & 49.05         & 34.56  \\
  \end{tabular}
  \caption{Normal mixture example: Average number of simulation steps per accepted particle for study with $N=5,000$ particles.} \label{table:tabToy}
\end{center}
\end{table}
\begin{figure}[h!]
\begin{center}
\includegraphics{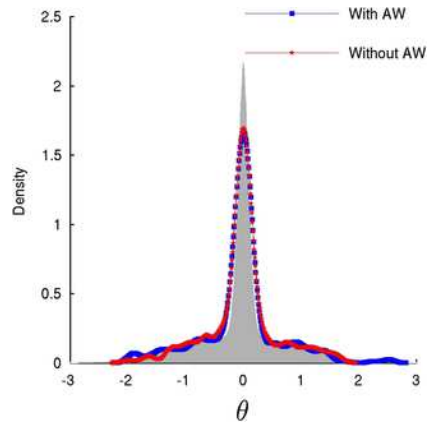}
\end{center}
\caption{Normal mixture example: Approximate (lines) and exact (shaded) posterior densities.}
\label{fig:figToy}
\end{figure}

\begin{figure}[htp!]
  \centering
\includegraphics{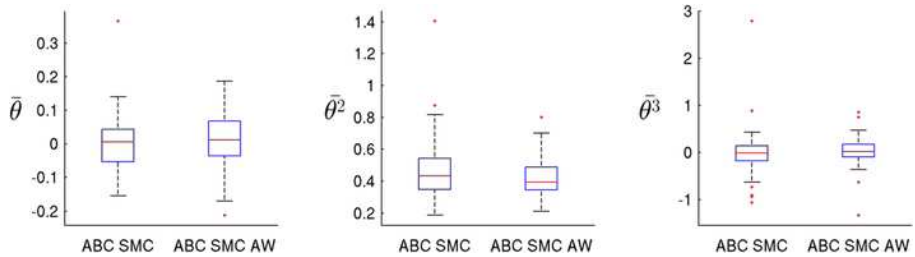}
\caption{\label{fig:figBoxplots_toy} Monte Carlo estimates for the normal mixture example: Box-plots of estimated posterior moments (as annotated) based on 50 repeated runs of SMC ABC and SMC ABC AW.}
\end{figure}

We ran ABC SMC and ABC SMC AW to obtain samples of $N=$5,000 particles in each case.
Table~\ref{table:tabToy} shows the average number of simulation steps per accepted particle; the adaptive weight
modification  requires about 30\% fewer simulations. As shown in Figure~\ref{fig:figToy}, the resulting posterior approximations
are  quite accurate.  More importantly from the viewpoint of the methodology here, the two strategies give very closely similar
results, with the AW variant substantially improving the computational efficiency in terms of acceptance rates.  The suggested equivalence of accuracy of the methods is also confirmed in Figure~\ref{fig:figBoxplots_toy}, which displays a comparison of Monte Carlo estimates based on 50 repeated runs of both methods using $N=$1,000 particles. The plots in the figure do not indicate any significant difference in the distribution of Monte Carlo estimates produced by both methods.
In terms of coefficient of variation of the final importance weights, the outputs of both methods were also similar, with average coefficient of 1.34 (with standard deviation 0.46) for ABC SMC AW, and average coefficient of 1.13 (with standard deviation 0.37) for ABC SMC.

\subsection{Multivariate Mixture Examples}
\label{sec:Normal_toy_multivariate}

To more aggressively explore performance, we have run studies on multivariate versions of the normal mixture example. Generally,
take $p-$dimensional data $\mathbf{x} |  \boldsymbol{\theta} \sim 0.5 \mathcal{N}_{p}( \boldsymbol{\theta},I_p) + 0.5 \mathcal{N}_p( \boldsymbol{\theta}, 0.01 I_p)$, where $\mathbf{x} = (x_1, \ldots, x_p )'$ and $\boldsymbol{\theta} = (\theta_1, \ldots, \theta_p)'$.  The prior has $\boldsymbol{\theta}$ uniformly distributed on $[-10,10]^p$.

For the method implementation we use normal kernels $K_{\theta,t}$ and $K_{x,t}$ with diagonal variance matrices
and with scalar bandwidths defined via the standard rule-of-thumb~\citep{Silverman1986,Scott2005} as used in the
example of the previous section. That is, for each scalar dimension $k$ of parameters and data,
if $\hat{\sigma}_k$ denotes the standard deviation in that
dimension, then  $h_k = \hat{\sigma}_k N^{-1/(d+4)}$ where $N$ is the number of particles and $d$ is the
total dimension (parameters and data). The discrepancy used is $\rho(x,\xo)=\sum_k (x_k - x_{k,obs})^2$.

Given the observation $\mathbf{x}_{obs} = (0,\ldots, 0)'$, we ran ABC SMC and ABC SMC AW to obtain samples of $N=$5,000 particles for different cases of dimension $p$. In each case, the threshold schedule was defined to keep correspondence with the schedule used in the example of Section~\ref{sec:Normal_toy}. This correspondence was achieved by selecting threshold values that result in the same percentiles of the discrepancy distribution of prior:model simulations.
The comparison results for repeat simulations for all cases with $p \in \{ 1,2,\ldots, 50 \}$ show substantial
reduction in the number of required  model simulations induced by the adaptive weight modification.
Across this range of dimensions, the reductions seen run from about $30-70\%$ (as seen in Table~\ref{table:tabToy_scale}),  confirming that the
adaptive weighting strategy can be effective in higher-dimensional problems, even with $p$ as high as 50 or more.

\begin{table}[h!]
  \begin{tabular}{ l c c c c c c c c c c c c }
dimension $p$       & 1             & 2            & 5         & 10         & 15         & 20         & 25         & 30         & 40         & 50       \\ \hline
ABC SMC         & 40.3  & 34.1  & 35.5  & 33.4  & 33.6  & 31.3  & 31.4  & 26.9  & 21.4  & 18.8 \\
ABC SMC AW      & 29.7      &14.4       & 10.2  & 11.4  & 12.8  & 13.1  & 13.9  & 12.2  & 11.4  & 11.4 \\
 \end{tabular}
  \caption{Multivariate normal mixture example: Average total number of simulation steps per accepted particle. } \label{table:tabToy_scale}
\end{table}



\section{Comparison in Applications}

Two studies demonstrate the improvements achievable using adaptive weights in a couple of interesting applied
contexts, one using real data and the other synthetic data (so that ``truth'' is known).
In each we use the same algorithm setup as described in Section~\ref{sec:Normal_toy_multivariate}.
For the threshold schedule, we
follow~\cite{Beaumont2002} with a pilot study of prior:model simulations to identify $\epsilon_T$ as a
very low percentile of the distribution of the discrepancies, and then define a schedule $\epsilon_{1:T}$ to
gradually reduce to that level. See also~\cite{Bonassi2011} for more insight into this approach and variants.
Specific values of the $\epsilon_t$  are noted in the following sections.

\subsection{Toggle Switch Model in Dynamic Bionetworks}

A real data example from systems biology comes from studies of dynamic cellular networks
based on measures of expression of genes (network nodes) at one or more
\lq\lq snapshots'' in time. The model, context and flow cytometry data set come from~\cite{Bonassi2011}. This
is data from experiments on bacterial cells using an engineered gene circuit with design related to the toggle switch
model~\citep{Gardner2000}. The model describes the dynamical behavior of a network with two genes ($u$ and $v$)
with doubly repressive interactions. In discrete time the specific model form is
\begin{equation}\begin{array}{lcl}
u_{c,t+h} & =  & u_{c,t} + h\alpha_u /(1+{v_{c,t}}^{\beta_u}) -h(1+ 0.03 u_{c,t})  + h 0.5 \xi_{c,u,t}, \\
v_{c,t+h} & =  & v_{c,t} + h\alpha_v/(1+{u_{c,t}}^{\beta_v})  -h(1+ 0.03 v_{c,t}) + h 0.5 \xi_{c,v,t},\\
\end{array}\label{eq:togglemodelstate}\end{equation}
over time $t,$ where $c$ indexes bacterial cells, $h$ is a small time step, the $\xi_{\cdot,\cdot,t}$ are independent
standard normals, and for some specified initial values $u_{c,0},v_{c,0}.$

The observable data set is a sample from the marginal distribution of levels of just one of the two genes at one point in time, viz.
$y = \{ y_c, c=1\mathord{:}\textrm{2,000}\},$ where $y_c$ is a noisy  measurement of $u_\tau$ at a given time point $\tau.$  The measurement error process is, cell-by-cell, given by
\begin{equation}
y_c \,\,\, = \,\,\, u_{c,\tau} +  \mu  +  \mu\sigma \eta_c/u_{c,\tau}^\gamma, \quad c=1,\ldots,\textrm{2,000},
\label{eq:togglemodeldata}\end{equation}
where the $\eta_c$ are standard normals, independent over cells and independent of the stochastic terms $\xi_{\cdot,\cdot,t}$  in the
state evolution model of equation~(\ref{eq:togglemodelstate}).
The study of~\cite{Bonassi2011} uses $h=1$, $\tau= 300$ and initial state $u_{c,0}=v_{c,0}=10$, adopted here.
The full set of 7 model parameters is $\theta=(\alpha_u,\alpha_v;\, \beta_u,\beta_v;\,  \mu,\sigma,\gamma).$

\begin{figure}[t!]
  \centering
\includegraphics{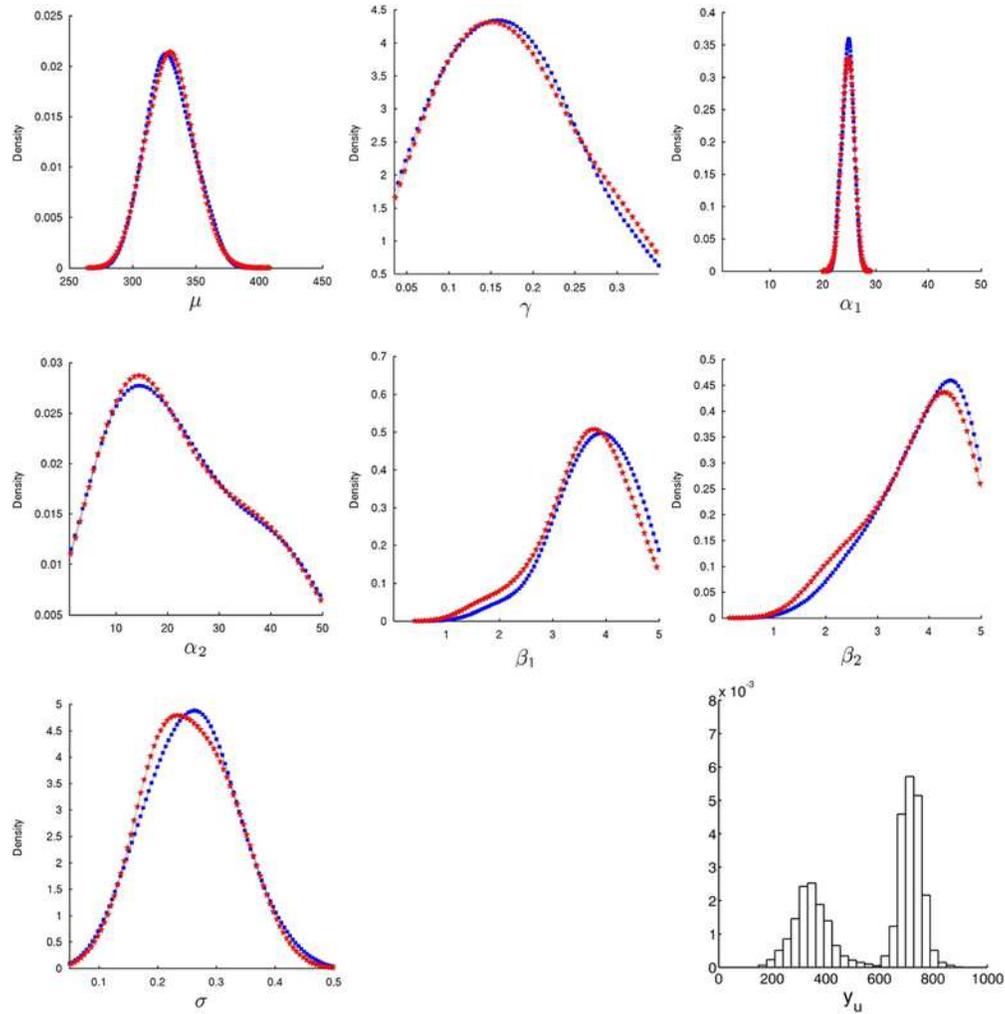}
\caption{\label{fig:figToggle} Toggle switch study:  Data are shown in the lower right frame (sample size $C=$2,000 cells).
The other frames represent approximate posterior marginal densities for each of the 7 toggle switch model
parameters, as annotated, to compare results from ABC SMC (red) and ABC SMC AW (blue).}\vspace*{-2pt}
\end{figure}

The data is shown in Figure~\ref{fig:figToggle}. Following~\cite{Bonassi2011}, the sample data set is reduced to a set of
summary \lq\lq reference signatures''  that define the effective data $x$ we aim to condition on.   The details of the
dimension reduction from the full, original sample $y$ to $x$ are not of primary interest here,  though they are of course
critically  important for the application area; readers interested in the specifics
of the applied context and the precise definition of $x$ can consult~\cite{Bonassi2011}. Here we simply start with the reduced, 11-dimensional data summary $x$, giving supplementary
code and data that provide the relevant details and precise definition of $x$. We do note that, in addition to dimension reduction, the specific definition of $x$ in~\cite{Bonassi2011}  has the advantage of producing an
orthogonal projection of transformed raw data so that the sample elements of $x$  are uncorrelated, which makes
our use of diagonal variance matrices in the kernels $K_{x,t}$ particularly apt.

\begin{table}[h!]
\begin{center}
  \begin{tabular}{ c c c c }
 $t$    & $\epsilon_t $ & ABC SMC & ABC SMC AW  \\
1   & 2500  & 12.3  & 12.2  \\
2   & 750   & 16.1  & 10.6  \\
3   & 250   & 24.3  & 20.6  \\
4   & 150   & 42.1  & 28.2  \\
5   & 75        & 96.4  & 61.8  \\
    \hline
    & Total     & 191.3     & 133.4 \\
  \end{tabular}
  \caption{Toggle switch study: Average number of simulations  per accepted particle  for study with $N=2,000$ particles.} \label{table:tabToggle}
\vspace*{-12pt}
\end{center}
\end{table}

The prior for $\theta$ is comprised of independent uniforms on real-valued transformed parameters, defined by finite ranges for
each based on substantive biochemical background information.\vadjust{\goodbreak} The priors are also taken from the prior study in~\cite{Bonassi2011}
and are  indicated by the ranges of the plots in Figure~\ref{fig:figToggle}.  We summarize  analyses with $N=$2,000 particles and
$\epsilon_{1:5} = (2500,750,250,150,75)$ where the final tolerance level $\epsilon_T=75$ corresponds to an approximate
$10^{-4}$ quantile of simulated prior discrepancies. Table~\ref{table:tabToggle} shows the average number of prior:data generation steps
per acceptance for both ABC SMC and ABC SMC AW.

Evidently, SMC ABC AW leads to a reduction of roughly 30\% simulation steps, a practically very material
gain in efficiency. Combined with this, Figure~\ref{fig:figToggle} shows that the approximate posteriors
from ABC SMC AW are practically the same as those from ABC SMC, the differences being easily attributable to
Monte Carlo variation.

The toggle switch  example is also used to illustrate some aspects of computation time of ABC SMC in a real application. We implemented the methods discussed by using a PC Intel 3.33 GHz. The total processing time for the ABC SMC algorithm was 70,543 seconds, while for the version with adaptive weights the time was 49,332 seconds. From that comparison we can observe the same gain in efficiency as observed in the number of required simulation steps, confirming the point that the computation of the adaptive weights adds negligible processing time to the ABC SMC
algorithm.

We also implemented the method of \cite{Delmoral2011} to illustrate the performance difference that is induced by the computation of the weights based on linear complexity in the number of particles. This implementation was based on the same settings discussed before, i.e., with the same number of particles, threshold schedule and normal kernel. The remaining element to be defined in that method is the number $M$ of synthetic data sets generated for each particle, and here we use $M=31$ so that the method relies on roughly the same overall number of simulations steps that were used in the ABC SMC implementation discussed before (without adaptive weights). Given this setup, the implementation of the method of \cite{Delmoral2011} was only 0.8\% faster than the previous ABC SMC implementation. This result confirms the point discussed in Section~\ref{sec:intro}, that the linear complexity in the computation of the weights does not necessarily result in significant gain in computation time for practical applications of ABC SMC.
\cite{Lenormand2012} present a more complete comparison of these methods under different scenarios, which suggests that the algorithm with linear complexity in the computation of the weights does not necessarily result in better overall computational efficiency. That study also takes into account the accuracy of the methods in the overall comparison.

\begin{figure}[t!]
  \centering
\includegraphics{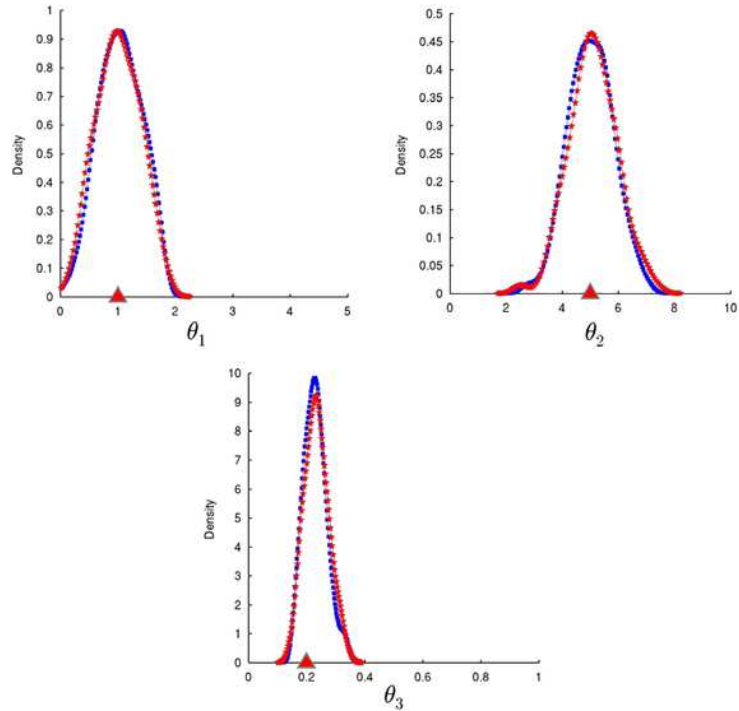}
\caption{\label{fig:figQueue} Queuing system analysis: Summaries of analyses of simulated data set where
the 3 frames represent margins for each of the model
parameters as annotated, showing results from ABC SMC (red) and ABC SMC AW (blue).  The true known parameter values
underlying the synthetic data are marked as red triangles on the horizontal axes. }
\end{figure}

\subsection{Queuing System}

The second study is of a queuing system previously discussed in~\cite{heggland2004} and addressed using
ABC by~\cite{Blum2010} and \cite{Fearnhead2011}. The context is a single server,  first-come-first-serve queue (M/G/1).
Three parameters $\theta=(\theta_1,\theta_2,\theta_3)'$ determine the distributions of service and inter-arrival  times; service times
are uniform on $[\theta_1,\theta_2]$ while inter-arrival times are  exponential  with rate $\theta_3.$ In the  notation of~\cite{heggland2004},
$W_r$ is the inter-arrival time of the $r$th customer and $U_r$ the corresponding service time. The inter-departure
time process \{$Y_r, r=1,2,\ldots$ \} is then
\begin{equation}
  Y_r = \left\{
  \begin{array}{l l}
    U_r, & \quad \text{if \, $\sum_{i=1}^r W_i  \le \sum_{i=1}^{r-1} Y_i $},\\
    U_r + \sum_{i=1}^r W_i - \sum_{i=1}^{r-1} Y_i& \quad \text{if \,  $\sum_{i=1}^r W_i  > \sum_{i=1}^{r-1} Y_i $}.\\
  \end{array} \right.
\end{equation}
Inter-arrival times are unobserved; the observed data from $R$ customers is \linebreak
$x= \{Y_1,\ldots,Y_R\}$ where $R$ is the number of customers.

We generated a synthetic data set following the specification in~\cite{Blum2010}; here $R=50$ with \lq\lq true'' parameters
$\theta=(1,5,0.2)'$. Summary statistics are taken as 3 equidistant quantiles together with the minimum and maximum values
of the inter-departure times. The prior has $(\theta_1, \theta_2 - \theta_1, \theta_3)$ uniformly distributed on $[0,10]^3$.

\begin{figure}[t!]
  \centering
\includegraphics{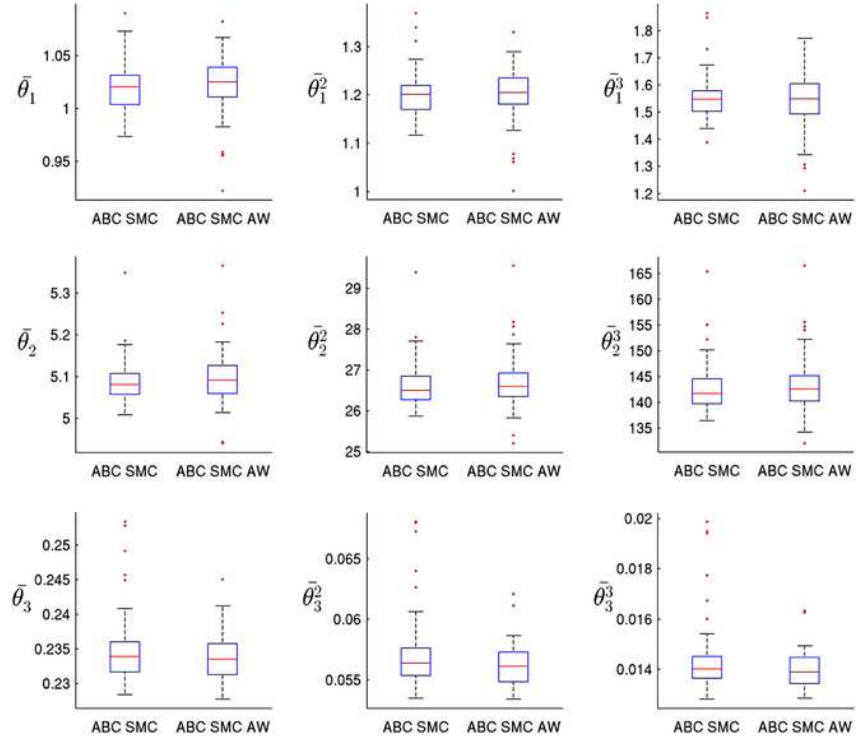}
\caption{\label{fig:figBoxplots_queue} Monte Carlo estimates for the queueing system example: Box-plots of estimated posterior moments (as annotated) based on 50 repeated runs of SMC ABC and SMC ABC AW.}
\end{figure}

Analysis used $N=$1,000 particles and discrepancy schedule $\epsilon_{1:5} = (200,100,10,2,1)$. The final level $\epsilon_T=1$
corresponds to a value close to the $10^{-4}$ quantile of simulated prior discrepancies.
Figure~\ref{fig:figQueue} shows close agreement between the estimated marginal posterior densities under ABC SMC and ABC SMC AW,
and, incidentally, that they support regions containing the true parameters underlying this synthetic data set.   Figure~\ref{fig:figBoxplots_queue} displays the comparison of Monte Carlo estimates based on 50 repeat runs of
the two methods using $N=$1,000 particles. The plots in the figure do not suggest any significant difference in the distribution of Monte Carlo estimates that result.
This similarity between the methods was observed for the coefficients of variation of the final importance weights. An average coefficient of 1.31 (with standard deviation 0.73) was observed for ABC SMC AW, and average coefficient of 0.92 (with standard deviation 0.65) for ABC SMC.

In order to study the gain in efficiency derived from the use of adaptive weights, we repeated this analysis 100 times, each repeat involving a new prior:data simulation. Table~\ref{table:tabQueue} shows summaries of the numbers of data generation steps necessary for the ABC SMC and
ABC SMC AW algorithms.   We see that we realized  an average reduction of about 58\% in the number of simulation steps by using adaptive weighting.

\begin{table}[h!]
\begin{center}
  \begin{tabular}{ c c c c c c c c c }
\   &  \ & \multicolumn{3}{c}{ABC SMC} & \ & \multicolumn{3}{c}{ABC SMC AW}  \\
\cline{3-5} \cline{7-9}
$t$    &    $\epsilon_t $ & min  & mean & max & \ & min  & mean & max \\
1   & 200   & 1.0 & 1.3  & 3.1 & \ & 1.0    & 1.3 & 3.0 \\
2   & 100   & 1.2 & 1.4  & 2.1 & \ & 1.0    & 1.0 & 1.4     \\
3   & 10    & 3.3 & 12.7 & 760.5 & \ & 1.2  & 2.4 & 60.6    \\
4   & 2            & 2.4 & 9.0  & 134.1  & \ & 1.4  & 3.9 & 39.9    \\
5   & 1         & 1.9 & 6.9  & 107.5 & \ & 1.1  & 4.5 & 80.7    \\ \hline
    & Total     &       & 31.3 &      & \  &       & 13.1 &    \\
  \end{tabular}
  \caption{Queuing system analysis: Summaries of 100 replicate synthetic data analyses, showing average numbers of simulation steps per accepted particle.} \label{table:tabQueue}
\end{center}
\end{table}

\section{Additional Discussion \label{sec:Additional Discussion}}

This paper has introduced ABC SMC AW, shown that it is theoretically expected to improve the effectiveness of
ABC SMC based on adaptive, data-based weights, and demonstrated some of the practical potential in two
interesting model contexts from related literature. The new approach
is simple to implement, requiring only a minor extension of standard ABC SMC code. Further, the computational
overheads adaptive weighting generates are-- in anything but trivial models-- typically quite negligible relative to the main expense of forward simulations of prior predictive distributions.
We also note that the adaptive weighting idea has the potential to be integrated into other ABC SMC extensions, even though this integration requires careful study of potential computational and theoretical implications.

The basic idea of adaptive weights links closely to  adaptive importance sampling and direct
posterior approximations based on mixtures of kernel forms.  As noted in Section~\ref{sec:AW}, the
local smoothing for nonlinear regression adjustment in ABC of~\cite{Bonassi2011} uses multivariate normal
mixtures in related ways as \lq\lq local" posterior approximations in regions defined by the threshold setting.
There the mixture modelling, used to define posterior emulators, is based on large-scale Bayesian nonparametric
models that can have many mixture components and so flexibly adapt to the shapes of local posterior contours.
Incidentally, mixture fitting is computationally effective based on GPU parallelized code~\citep{Suchard2010,Cron2010}.
This connection suggests a more general adaptive weighting strategy that uses a joint kernel that is
not of product form, with the potential to customize the weighting of sampled parameters further.
That is, in the AW algorithm of Figure~\ref{fig:ABC_SMC_AW}, the kernel
$K_{\theta,t}(\theta | \theta^*)$ would be modified to have shape characteristics that depend also
on the locale in which the kernel location $\theta^*$ sits. Building on the connections with multivariate
normal mixtures suggests specific ways in which these modifications could be developed, and this
is under investigation.  At this point, however, it is unclear just how beneficial this will be in practice.
Such extensions will require substantial additional computational overheads to identify and compute local
kernel functions, which may more than offset the gains in efficiency the extensions can be expected to generate.
The gains in acceptance rates already achieved by the very simple (to code and run) adaptive weighting
method of this paper can already be very substantial, as our examples highlight.

\bibliographystyle{ba}

%

\begin{acknowledgement}
This work was supported in part by grants from the U.S. National Science Foundation (DMS-1106516)
and National Institutes of Health (P50-GM081883 and RC1-AI086032).
Any opinions, findings and conclusions or recommendations expressed in this work are
those of the authors and do not necessarily reflect the views of the NSF or NIH.
\end{acknowledgement}

\end{document}